\documentclass[manuscript,screen]{acmart}

\AtBeginDocument{%
  \providecommand\BibTeX{{%
    \normalfont B\kern-0.5em{\scshape i\kern-0.25em b}\kern-0.8em\TeX}}}

\setcopyright{acmcopyright}
\copyrightyear{2020}
\acmYear{2020}
\acmDOI{xxx/yyy}

\usepackage{comment}
\usepackage{mathtools}
\usepackage{booktabs}

\DeclarePairedDelimiter\abs{\lvert}{\rvert}%
\begin{document}

\title{Experience: Improving Opinion Spam Detection by Cumulative Relative Frequency Distribution}

\author{Michela Fazzolari}
\affiliation{%
  \institution{Istituto di Informatica e Telematica, Consiglio Nazionale delle Ricerche}
  \streetaddress{Via G. Moruzzi, 1}
  \city{Pisa}
  \country{Italy}}
\email{m.fazzolari@iit.cnr.it}
\orcid{0000-0002-8562-6904}

\author{Francesco Buccafurri}
\affiliation{%
  \institution{DIIES, Universit\`{a} Mediterranea di Reggio Calabria}
  \streetaddress{Via Graziella, Localit\`a Feo di Vito}
  \city{Reggio Calabria}
  \country{Italy}}
\email{bucca@unirc.it}
\orcid{0000-0003-0448-8464}

\author{Gianluca Lax}
\affiliation{%
  \institution{DIIES, Universit\`{a} Mediterranea di Reggio Calabria}
  \streetaddress{Via Graziella, Localit\`a Feo di Vito}
  \city{Reggio Calabria}
  \country{Italy}}
\email{lax@unirc.it}
\orcid{0000-0002-5226-0870}

\author{Marinella Petrocchi}
\affiliation{%
  \institution{Istituto di Informatica e Telematica, Consiglio Nazionale delle Ricerche}
  \streetaddress{Via G. Moruzzi, 1}
  \city{Pisa}
  \country{Italy}}
\email{m.petrocchi@iit.cnr.it}
\orcid{0000-0003-0591-877X}

\renewcommand{\shortauthors}{Fazzolari et al.}

\begin{abstract}
Over the last years, online reviews became very important since they can influence the purchase
decision of consumers and the reputation of businesses, therefore, the practice of writing fake reviews can have severe consequences on customers and service providers. Various approaches have been proposed for detecting opinion spam in online reviews, especially based on supervised classifiers. In this contribution, we start from a set of effective features used for classifying opinion spam and we re-engineered them, by considering the Cumulative Relative Frequency Distribution of each feature. 
By an experimental evaluation carried out on real data from Yelp.com, we show that the use of the distributional features is able to improve the performances of classifiers.
\end{abstract}



\maketitle
\section{Introduction}
As illustrated in a recent survey on the history of Digital Spam~\citep{Ferrara2019cacm}, the Social Web  has led not only to a `participatory, interactive nature of the Web experience', but also to the proliferation of new and widespread forms of spam, among which the most notorious ones are fake news and spam reviews, {\it a.k.a.} opinion spam. This results in the diffusion of different kinds of disinformation and misinformation, where misinformation refers to inaccuracies that may even originate acting in good faith, while disinformation is false information deliberately spread to deceive~\citep{hernon1995disinformation}.

Over the last years, online reviews became very important since they reflect the customers' experience with a product or service and, nowadays, they constitute the basis on which the reputation of an organization is built. 
Unfortunately, the confidence in such reviews is often misplaced, due to the fact that spammers are tempted to write fake information in exchange for some reward or to mislead consumers for obtaining business advantages~\citep{Liu2012sentiment}.

The practice of writing false reviews is not only morally deplorable, as it is misleading for customers and inconvenient for service providers, but it is also punishable by law. 
Considering both the longevity and the spread of the phenomenon, scholars for years have investigated  various approaches to opinion spam detection, 
mainly based on supervised or unsupervised learning algorithms. Further approaches are based on Multi Criteria Decision Making~\citep{PasiVC19}.

Machine learning approaches rely on input data to build a mathematical model in order to make predictions or decisions. To this aim, data are usually represented by a set of features, which are structured and ideally fully representative of the phenomenon being modeled. An effective feature engineering process, i.e., the process through which an analyst uses the domain knowledge of the data under investigation to prepare appropriate features~\citep{Crawford2015survey}, is a critical and time-consuming task. 
However, if done correctly, feature engineering increases the predictive power of algorithms by facilitating the machine learning process.

In this paper, we do not aim to contribute by defining novel features suitable for fake reviews detection, 
rather, starting from features that have been proven to be very effective by Academia, we {\it re-engineered} them, by considering the distribution of the occurrence of the features values in the dataset under analysis.  
In particular, we focus on the Cumulative Relative Frequency Distribution of a set of the basic features already employed for the task of fake review detection. We compute this distribution for each feature and substitute each feature value with the corresponding value of the distribution. To demonstrate the effectiveness of the proposed approach, the \textit{distributional (cumulative) features} and the \textit{basic} ones have been exploited to train several supervised machine-learning classifiers and the obtained results have been compared. To the best of the authors' knowledge, this is the first time that Cumulative Relative Frequency Distribution of a set of features has been considered for the unveiling of fake reviews. 
The experimental results show that the distributional features improve the performances of the classifiers, at the mere cost of a small computational surplus in the feature engineering phase.

The rest of the paper is organized as follows. The next section revises related work in the area. Section~\ref{sec:features} describes the process of feature engineering. In Section~\ref{sec:setup}, we present the experimental setup, while Section~\ref{sec:results} reports the results of the comparison among the classification algorithms. Moreover, in this section, we assess the importance of the distributional features and discuss about the benefits brought by their adoption. Finally, Section~\ref{sec:concl} concludes the paper.

\section{Related Work}
\label{sec:RW}
Social Media represent the perfect means for everyone to ``spread contents  in
the form of User-Generated Content (UGG), almost without any traditional form of trusted control''~\cite{abs-2001-09473}.
Since years, Academia, Industry, and Platform Administrators have been fighting for developing automatic solutions to raise the users' awareness about the credibility of the news they read online.
One of the contexts in which the problem of credibility assessment
is receiving the most interest is spam - or fake - reviews detection. The existence of spam reviews has been known since the early 2000s when e-commerce and e-advice sites began to be popular.

In his seminal work, Liu lists three approaches to automatically identify opinion spam: the supervised, unsupervised, and group approaches~\cite{Liu2012sentiment}. In a standard supervised approach, a ground truth of a priori known genuine and fake reviews is needed. Then, features about the labeled reviews, the reviewers, and the reviewed products are engineered. The performances of the first models built on such features achieved good results with common algorithms such as Naive Bayes and Support Vector Machines~\cite{mukherjee2013spotting}. 

As usual, a supervised approach is particularly challenging since it requires the existence of labeled data, that is, in our scenario, a set of reviews with prior knowledge about their (un)trustworthiness. To overcome the frequent issue of lack of labeled data, in the very first phases of investigation in this field, the work done by Jindal et al. in~\cite{Jindal2008opinion} exploited the fact that a common practice of fraudulent reviewers was to post almost duplicate reviews: reviews with similar texts were collected as fake instances. As shown in ~\cite{Crawford2015survey}, linguistic features have been proven to be valid for fake reviews detection, particularly in the early advent of this phenomenon. Indeed, pioneer fake reviewers exhibited precise stylistic features in their texts, such as a marked use of short terms and expressions of positive feelings. Anomaly detection was also been widely employed in this field: an analysis of anomalous practices with respect to the average behavior of a genuine reviewer led to good results. Anomalous behavior of the reviewer may be related to general and early rating deviation, as highlighted by Liu in~\cite{Liu2012sentiment}, or temporal dynamics (see Xie et al.~\cite{Xie:2012}).


Going further with the useful methodologies, human annotators, possibly recruited from crowd-sourcing services like  Amazon Mechanical Turk~\cite{AMT}, have also been employed, both 1) to manually label reviews' sets to separate  fake from non-fake reviews (e.g., see the very recent survey by Crawford et al. in~\cite{Crawford2015survey}) and 2) to let them write intentionally false reviews, in order to test the accuracy of existing predictive models on such set of ad hoc crafted reviews, as nicely reproduced by Ott et al. in~\cite{Ott2011deceptive}. 

Recently, an interesting point of view has been offered by Cocarascu and Tonotti in~\cite{cocarascu2018combining}: deception is analysed based on contextual information derivable from review texts, 
but not  in a standard way, e.g., considering linguistic features, but evaluating the influence and interactions that one text has on the others. The new feature, based on bipolar argumentation on the same review, has been shown to outperform more traditional features, when used in standard supervised classifiers, and even on small datasets. 

Supervised learning algorithms usually need diverse examples - and the values of diverse features derived from such examples - for an accurate training phase. Wang et al. investigated the `cold-start' problem~\cite{wang2017handling}: the identification of a fake review when a new reviewer posts one review. Without enough data about the stylistic features of the review and the behavioral characteristics of the reviewer, the authors first
find similarities between the review text under investigation and other review texts.  Then, they consider similar behavior between the reviewer under investigation and the reviewers who posted the identified reviews. A model based on neural networks proves to be effective to approach the problem of lack of data in cold-start scenarios.

Although many years have passed and, as we will see briefly later, the problem has been addressed in many research works, with different techniques,  automatically detecting a false review is an issue not completely solved yet,
as stated in the recent survey of Wu et al.~\cite{wu2020fake}. 
This inspiring work examines the phenomenon not only giving an overview of the various detection techniques used over time, but also proposing twenty future research questions. Notably, to help scholars find suitable datasets for a supervised classification task, this survey lists the currently available review datasets and their characteristics.

A similar work by Hussain et al~\cite{hussain2019detecting}, aimed at a comparison of different approaches, focuses on the performances obtained by different classification frameworks. Also, the authors carried on a relevance analysis of six different behavioral features of reviewers. Weighting the features with respect to their relevance, a classification over a baseline dataset obtains an 84.5\% accuracy. 

A quite novel work considers the unveiling of malicious reviewers by exploiting the notion of `neighborhood of suspiciousness'. In~\cite{squicciarini2018}, Kaghazgaran et al. proposed a system called TwoFace that, starting from identifiable reviewers paid by crowd-sourcing platforms to write fake reviews on well-known e-commerce platforms, such as Amazon, studies the similarity between these and other reviewers, based, e.g., on the reviewed products, and shows how it is possible to spot organized fake reviews campaigns  even when the reviewers alternate  genuine and malicious behaviors. 

Serra et al. developed a supervised approach where the task is to differentiate amongst different kinds of reviewers, from fraudulent, to uninformative, to reliable~\cite{SerraSSS20}. Leveraging a supervised classification approach based on a deep recurrent neural network, the system achieves notable performances over a real dataset where there is an a priori knowledge of the fraudulent reviewers. 

The research work reminded so far lies in supervised learning. However, unsupervised techniques have been employed too, since they are very useful when no tagged data is available.
As an example, the authors of~\cite{dhingra2019spam} 
start from the same hypothesis as the above-cited~\cite{squicciarini2018} on the classification of the kind of reviewers. A reviewer may not always be considered either fraudulent or honest. Indeed, a behavioral analysis of the reviewer may leave a degree of uncertainty that can lead to classification errors. So, the work in~\cite{dhingra2019spam} proposed 
an unsupervised learning approach based on fuzzy logic, developing a new deductive algorithm able to obtain about 80\% accuracy on the classification of a group of reviewers, as belonging to one of seven categories of suspiciousness. 

Lack of annotated datasets is not the only reason to resort to other than a supervised approach. In fact, the quality of available datasets can be questionable. The datasets need to be constantly updated, and it has been demonstrated that a human-operated annotation process is prone to error~\cite{Ott2011deceptive}. In this regard, the work by Rout et al. in~\cite{rout2017revisiting} investigated semi-supervised approaches, i.e., approaches in which a small amount of labeled data is used in combination with a large amount of unlabeled data during training. The adoption of four well-known semi-supervised learning approaches shows a promising increase in the classification performances.

In recent years, a behavioral analysis of the target under investigation has been proven to be useful not only to discover individual fake users, but also to detect the coordinated and synchronized behavior that characterizes groups of  malicious users. For some years now, acting in groups is a very common practice that characterizes social bots, i.e., automated social accounts programmed for unethical and often illicit online trafficking~\cite{abs-2007-03604,cresci2019capability,CresciPST19}. They have been massively employed to influence and alter the public opinion in major events, such as political elections and societal debates (e.g.,  see Badawi et al. about the 2016 US presidential elections~\cite{BadawyALF19} and Caldarelli et al. about the immigration from North Africa to Italy~\cite{caldarelli2020role}). Also in the field of electronic word of mouth, some researchers have highlighted how it is possible to find reviewers, in this case real humans, paid to review the same product with predefined schemes and timing\footnote{`Why I write fake online reviews', Online: https://www.bbc.com/news/uk-47952165}.
Fake reviewers' coordination can emerge by mining frequent behavioral patterns and ranking the most suspicious ones. A pioneer work by ~\cite{MukherjeeLG12} first identifies groups of reviewers that reviewed the same set of products; then, the authors compute and aggregate an ensemble of anomaly scores (e.g., based on similarity amongst reviews and times at which the reviews have been posted): the scores are ultimately used to tag the reviewers as colluding or not. 
Another interesting approach for the analysis of colluding users is the one proposed by~\cite{viswanath2015}: the authors check whether a given group of accounts (e.g., reviewers on {\it Yelp}) contains a subset of malicious accounts.
The intuition behind this methodology is that the statistical distribution of reputation scores (e.g., number of friends and followers) of the accounts participating in a tampered computation significantly diverges from that of untampered ones.

We close this section by referring back to the division made by Liu in~\cite{Liu2012sentiment} about supervised, unsupervised, and group approaches to spot fake reviewers and/or reviews. As noted in \cite{abs-2001-09473}, these are {\it data-driven} classification methods, mostly aiming at classifying in `a binary or multiple
way information items (i.e., credible vs non-credible)' with the evaluation of a  series of credibility features extracted from the data. Notably, {\it model-driven} approaches, which are based on some prior domain knowledge, are promising in providing a ranking of the information item (i.e., in our scenario, of the review) with respect to credibility. This is the case of recent work by Pasi et al.~\cite{PasiVC19,VivianiP17}, which 
exploits a Multi-Criteria Decision Making approach to assess the credibility of a review. In this context, a given review, seen as an alternative amongst others, is evaluated with respect to some credibility criteria. An overall credibility estimate of the review is then obtained by means of a suitable model-driven approach based on aggregation operators. This approach has also the advantage of assessing the contribution that single or interacting criteria/features have in the final ranking. 

The techniques presented above have their pros and cons, and depending on the particular context, one approach can be preferred with respect to another. 
The most relevant contribution of our proposal with respect to the state of the art is to improve the effectiveness of the solution based on supervised classifiers, which, as seen above, is a well-known and widely-used approach in this context.

\section{Feature Engineering}\label{sec:features}
In this section, we introduce 
a subset of features that have been adopted in past work to detect opinion spam and we propose how to modify them in order to improve the performances of classifiers. We emphasize that the listed features have been used effectively for this task by past researchers. We give below the rationale for their use in the context of unveiling fake reviews. Finally, it is worth noting that the list of selected features is not intended to be exhaustive.

\subsection{Basic Features}\label{sec:basic_features}

Following a supervised classification approach, the selection of the most appropriate features plays a crucial role, since they may considerably affect the performance of the machine learning models constructed starting from them~\citep{Crawford2015survey}.

Features can be review-centric or reviewer-centric~\citep{jindal2007analyzing}. The former are features that refer to the review, while the latter refer to the reviewer.
In the literature, several reviewer-centric features have been investigated, such as the maximum number of reviews, the percentage of positive reviews, the average review length, the reviewer rating deviation~\citep{mukherjee2013what}.
According to the outcomes of several works proposed in the context of opinion spam detection ~\citep{mukherjee2013what,mukherjee2013spotting,rayana2015collective}, we focused on reviewer-centric features, which have been demonstrated to be more effective for the identification of fake reviews.  Thus, we relied on a set of \textit{basic features}, which have been already used proficiently in the literature for the detection of opinion spam in reviews. Specifically, we focused on the following reviewer-centric features:
\begin{itemize}

\item \textbf{Photo Count}: This metric measures the number of pictures uploaded by a reviewer and is directly retrieved from the reviewer profile. In~\cite{zhang2016what}, the authors demonstrated the effectiveness of using photo count, together with other non-verbal features, for detecting fake reviews. 

\item \textbf{Review Count}: It measures how many reviews have been posted by a reviewer on the platform.  ~\cite{mukherjee2013spotting,mukherjee2013what} showed that spammers and non-spammers present different behavior regarding the number of reviews they post. In particular, spammers usually post more reviews, since they may get paid. This feature has also been investigated by~\cite{luca2013fake} and \cite{wang2013quantifying}.

\item \textbf{Useful Votes}: The most popular online review platforms allow users to rank reviews as useful or not. This information can be retrieved from the reviewer profile, or computed by summing the total amount of useful votes received by a reviewer. This feature has already been exploited by~\cite{zhang2016what} and it has been demonstrated to be effective for opinion spam detection. 
    
\item \textbf{Reviewer Expertise}: Past research in~\cite{mukherjee2013spotting,xu2013detecting} highlights that reviewers with acquired expertise on the platform are less prone to cheat. Particularly, Mukherjee et al. in~\cite{MukherjeeLG12} report that opinion spammers are usually not longtime members of a site. Genuine reviewers, however, use their accounts from time to time to post reviews. Although this experimental evidence does not mean that no spammer can be a member of a review platform for a long time, the literature has considered useful to exploit the activity freshness of an account in cheating detection. The Reviewer Expertise has been defined by Zhang et al. in ~\cite{zhang2016what} as the number of days a reviewer has been a member of the platform (the original name was Membership Length).
    
\item \textbf{Average Gap}: The review gap is the time elapsed between two consecutive reviews. 
This feature has been previously introduced in the seminal work by Mukherjee et al., under the name {\it Activity Window} ~\cite{mukherjee2013fake}, and successfully re-adopted for detecting both colluders (i.e., spammers acting with a coordinated strategy) ~\cite{mukherjee2013spotting,mukherjee2013what} and singleton reviewers (i.e., reviewers with just isolated behavioral posting) ~\cite{fei2013exploiting}. In the cited work, the Activity window feature as been proved highly discriminant for demarcate spammers and non-spammers. Quoting from~\cite{mukherjee2013fake}, ``fake reviewers are likely to review in short bursts and are usually not longtime active members''. On a Yelp dataset where was a priori known the benign and malicious nature of  reviewers, work in~\cite{mukherjee2013fake} proved that, by computing the difference of timestamps of the last and first reviews for all the reviewers, a majority (80\%) of spammers were bounded by 2 months of activity, whereas the same percentage of non-spammers remain active for at least 10 months.
We define the Average Gap feature as 
the average time, in days, elapsed between two consecutive reviews of the same reviewer and is defined as:
\begin{equation*}
   AG_i = \frac{1}{N_i - 1} \sum_{j=2}^{N_i} (T_{i,j} - T_{i,j-1})
\end{equation*}
where $AG_i$ is the Average Gap for the $i$-th user, $N_i$ is the number of reviews written by the user, $T_{i,j}$ is the timestamp of the $j$-th reviews of the $i$-th user. 

\item \textbf{Average Rating Deviation}: The rating deviation measures how much a reviewer's rating is far from the average rating of a business. \cite{lim2010detecting} observed that spammers are more prone to deviate from the average rating than genuine reviewers. However, a bad experience may induce a genuine reviewer to deviate from the mean rating. The Average Rating Deviation is defined as follows \citep{Jindal2008opinion,lim2010detecting,fei2013exploiting,mukherjee2013what}:
\begin{equation*}
    ARD_i = \frac{1}{N_i} \sum_{j=1}^{N_i} \abs{R_{i,j} - {R_{B(j)}}}
\end{equation*}
where $ARD_i$ is the Average Rating Deviation of the $i$-th user, $N_i$ is the number of reviews written by the user, $R_{i,j}$ is the rating given by the $i$-th user to her/his $j$-th reviews corresponding to the business $B(j)$, ${R_{B(j)}}$ is the average rating obtained by the business $B(j)$.

\item \textbf{First Review}: Spammers are usually paid to write reviews when a new product is placed on the market. This is due to the fact that early reviews have a great impact on consumers' opinions and, in turn, impact the sales, as pointed out by~\cite{lim2010detecting} and \cite{mukherjee2013what}. 
We compute the time elapsed between each review of a reviewer and the first review, for the same business. Then, we average the results on all the reviews. Specifically, the First Review value for reviewer $i$ is given by:
\begin{equation*}
        FRT_i = \frac{1}{N_i} \sum_{j=1}^{N_i} (T_{i,j} - F_{B(j)})
\end{equation*}
where $FR_i$ is the First Review value of the $i$-th user, $N_i$ is the number of reviews written by the user, $T_{i,j}$ is the time the $i$-th user wrote the $j$-th review and $F_B(j)$ is the time the first review of the same business $B(j)$, corresponding to the one of the $j$-th review, has been posted.

\item \textbf{Reviewer Activity}: Several works pointed out that the more active a user on the online platform, the more the user is likely genuine~\cite{goes2014popularity}, 
in terms of contributing with knowledge sharing in a useful way.
The usefulness of this feature has been demonstrated several years ago. Since the early 00s, surveys have been conducted on large communities of individuals, trying to understand what drives them to be active and useful on an online social platform, in terms of sharing content~\cite{wasko2005why}. Results showed that people contribute their knowledge when they perceive that it enhances their reputations, when they have the experience to share, and when they are structurally embedded in the network.

The Activity feature expresses the number of days a user has been active and it is computed as:
\begin{equation*}
    A_i = T_{i,L} - T_{i,0}
\end{equation*}
where $A_i$ is the activity (expressed in days) of the $i$-th user, $T_{i,L}$ is the time of the last review of the $i$-th user and $T_{i,0}$ is the time of the first review of the $i$-th user.
\end{itemize}

\subsection{From Basic to Cumulative Features}\label{sec:engineering}
The features described so far have been used to train a machine learning algorithm to construct a classifier, in a supervised-learning fashion. 

In this work, we propose to build on the basic features, with a proper feature engineering process, to possibly assess an improvement of the classification performances. The proposed feature engineering process is based on the concept of Cumulative Relative Frequency Distribution. 

The Relative Frequency is a quantity that expresses how often an event occurs divided by all outcomes. It can be easily represented by a Relative Frequency table. The Relative Frequency table is constructed directly from the data by simply dividing the frequency of a value by the total number of values in the dataset. 

The Cumulative Relative Frequency is then calculated by adding each frequency from the Relative Frequency table to the sum of its predecessors. In practice, the Cumulative Relative Frequency indicates the percentage of elements in the dataset that lies below the current value.
In this work, we modify each feature by using its Cumulative Relative Frequency Distribution. 

In the following, we show an example of how to compute the Cumulative Relative Frequency Distribution.
Let us consider the \textit{Photo Count} feature and assume that for each photo count value, the corresponding number of occurrences is the one reported in the second column of Table~\ref{tab:freq_table}. Thus, the second column reports the number of reviews associated with a reviewer who uploaded a given number of photos: in our example, there are 7,944 reviews whose reviewers have no photo associated. The third column measures the Relative Frequency, which is computed by dividing the number of occurrences by the total number of reviews. Finally, the fourth column reports the Cumulative Relative Frequency values, which have been obtained by adding each Relative Frequency value to the sum of its predecessor.

In our proposal, the process described so far is carried out for each basic feature and the Cumulative Relative Frequency values are used to train the classifier instead of the simple values. This involves, in practice, to substitute each value of the first column with the corresponding value of the fourth column of Table~\ref{tab:freq_table}.
\begin{table}[!bht]
    \centering
    \footnotesize
    \begin{tabular}{lrcc}
    \toprule
    \textbf{Photo} & \textbf{Frequency} & \textbf{Relative.Freq.} & \textbf{Cumulative} \\ 
    \textbf{Count} & \textbf{(\#Reviews)} & \textbf{(\%Reviews)} & \textbf{Rel.Freq.} \\ 
    \midrule
    0  &      7944 & 0.44 & 0.44 \\
    1  &      2301 & 0.13 & 0.57 \\
    2  &      1756 & 0.10 & 0.67 \\
    3  &      1401 & 0.08 & 0.75 \\
    4  &       822 & 0.04 & 0.79 \\
    5  &      1382 & 0.08 & 0.87 \\
    \bottomrule
    \end{tabular}
    \hfill
        \begin{tabular}{lrcc}
    \toprule
    \textbf{Photo} & \textbf{Frequency} & \textbf{Relative.Freq.} & \textbf{Cumulative} \\ 
    \textbf{Count} & \textbf{(\#Reviews)} & \textbf{(\%Reviews)} & \textbf{Rel.Freq.} \\ 
    \midrule
    6  &       550 & 0.03 & 0.90 \\
    7  &       347 & 0.02 & 0.92 \\
    8  &       780 & 0.04 & 0.96 \\
    9  &       342 & 0.02 & 0.98 \\
    10 &       460 & 0.02 & 1.00 \\
       &           &      &      \\
    \bottomrule
    \end{tabular}
    \caption{An example of Frequencies, Relative Frequencies, and Cumulative Relative Frequencies values for the Photo Count feature.}\label{tab:freq_table}
\end{table}

\section{Experimental Setup}
\label{sec:setup}
In this section, we describe the setting of the experiments conducted to evaluate the effectiveness of the proposed features. 
This is done by comparing the results obtained when using the basic features and the cumulative ones with the most widespread supervised machine learning algorithms. 

\subsection{Dataset Construction and Characteristics}\label{sec:datasetorigin}
The dataset used in this study is composed of 56,317 business reviews, 42,673 businesses (both restaurants and hotels), and 1,429 reviewers.
This dataset has been obtained by repopulating the YelpCHI dataset~\citep{YelpCHI}. The YelpCHI dataset included 67,395 reviews from 201 hotels and restaurants done by 38,063 reviewers and each review was tagged with a fake/non-fake label. To tag reviewers in the YelpCHI dataset and to obtain fresher data to work with, we operate as follows:

\begin{enumerate}
\item 
\textit{Reviewers tagging}. We assign a fake/non-fake label to each reviewer in the following way: if all the YelpCHI reviews of a single reviewer are tagged as fake (non-fake), then we consider the corresponding reviewer as fake (non-fake).
Instead, reviewers who presented a mix of fake and non-fake reviews in YelpCHI are tagged as \textit{mix}.  At the end of this elaboration, we measure in YelpCHI  79.7\% non-fake reviewers, 20\% fake reviewers, and 0.3\% \textit{mix} reviewers, which are discarded due to their limited number.
\item  \textit{Dataset repopulation}. For the fake and non-fake reviewers, we crawl again their reviews from the Yelp website. The updated dataset includes reviews from 2005 to September 2018.  Moreover, for each reviewer, we retrieve additional information, namely: 1) the number of posted photos, 2) the number of received useful votes, and 3) the date of registration to the platform.
Some of the reviewers were no longer available on the Yelp website and we could not update their profile, therefore they were discarded. 
\end{enumerate}

\begin{figure}[tb!]
    \centering
    \includegraphics[width=0.7\textwidth]{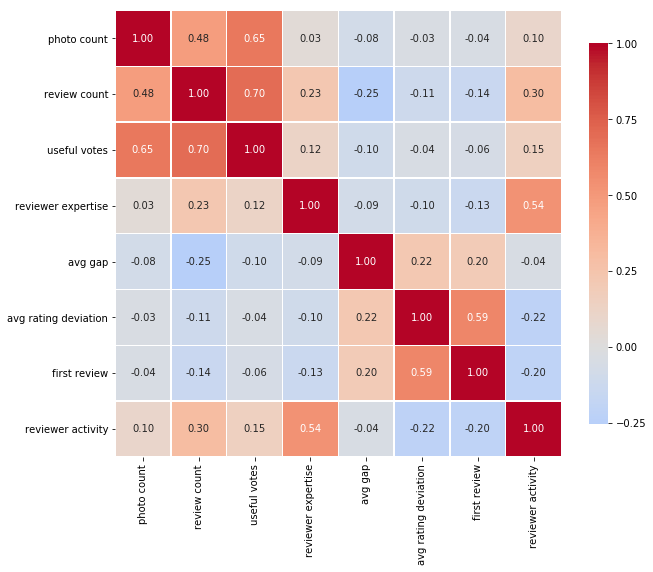}
    \caption{Correlation matrix showing the correlation coefficients among the basic features.}
    \label{fig:corr_mat}
\end{figure}

In Table~\ref{tab:dataset_summary} we report a summary with the statistics of the basic features for the repopulated dataset, whereas in Figure~\ref{fig:corr_mat} we present the correlation matrix, which shows the correlation coefficients among variables.

\begin{table}[!htb]
    \centering
    \begin{tabular}{lrrrrrrrr}
        \toprule
         & \textbf{photo} & \textbf{review} & \textbf{useful} & \textbf{reviewer} & \textbf{avg} & \textbf{avg rating} & \textbf{first} & \textbf{reviewer}  \\
         & \textbf{count} & \textbf{count} & \textbf{votes} & \textbf{expertise} & \textbf{gap} & \textbf{deviation} & \textbf{review} & \textbf{activity}  \\
         \midrule
        \textbf{mean} & 170.9 & 201.9 & 502.7 & 3664.7 & 55.2 & 0.01 & 13.9 & 2637.6  \\
        \textbf{std\_dev} & 911 & 298.2 & 2089.45 & 579.8 & 	110.6 &	0.06 & 	50.2 & 991.3  \\
         \bottomrule
         
    \end{tabular}
    \caption{Mean and standard deviation of the basic features for the repopulated dataset.}
    \label{tab:dataset_summary}
\end{table}

From the correlation matrix, we notice a slightly positive correlation between \textit{photo count} and \textit{review count}, between \textit{reviewer activity} and \textit{reviewer expertise}, between \textit{first review} and \textit{average rating deviation}, and between \textit{useful votes} and \textit{photo count}. A stronger positive correlation is highlighted between \textit{useful votes} and \textit{review count}.

\subsection{Data Labeling}
One limitation of supervised classification approaches is the possible lack of labeled data. To overcome this problem, \cite{ott2012estimating} relied on Amazon Mechanical Turks to generate fake and genuine reviews. Nevertheless, ~\cite{mukherjee2013what} highlighted the limits of this approach, since workers could be not always effective in imitating the behavior of fake reviewers, and proposes to use the results of the Yelp classification algorithm for obtaining labeled data. The same approach has also been used by~\citep{rahman2015tocatch} and ~\citep{zhou2016doprofessional} and the literature has recognized its effectiveness in detecting fake reviews~\citep{luca2013fake,zhang2016what}. 
According to this result, we use the same approach
to build a dataset of labeled examples 
(as described in Section~\ref{sec:datasetorigin}).

\subsection{Experimental Framework}\label{sec:exp_frame}
We experiment several supervised machine-learning algorithms, namely Logistic Regression (LogReg)~\citep{menard2002applied}, Linear Discriminant Analysis (LDA)~\citep{mclachlan2004discriminant}, Support Vector Machine (SVM)~\citep{cortes1995support}, Decision Tree (DT)~\citep{breiman1984classification}, Naive Bayes (NB)~\citep{maron1961automatic}, K-Nearest Neighbors (k-NN)~\citep{altman1992anintroduction}.
    
The performances of the learning algorithms are evaluated by means of commonly used metrics. Specifically, we compute the Accuracy on train (TraAcc) and test (TstAcc), the Precision (Pre), Recall (Rec) and F1-score (F1) for both classes. In addition, we compute the Matthews Correlation Coefficient (MCC)~\citep{matthews1975comparison} and the Area Under Curve (AUC)~\citep{ling2003auc}, which are two evaluation metrics used to measure the quality of binary classifications and are useful to compare classification models' performances.

\begin{figure}[!b]
    \centering
    \includegraphics[width=0.7\textwidth]{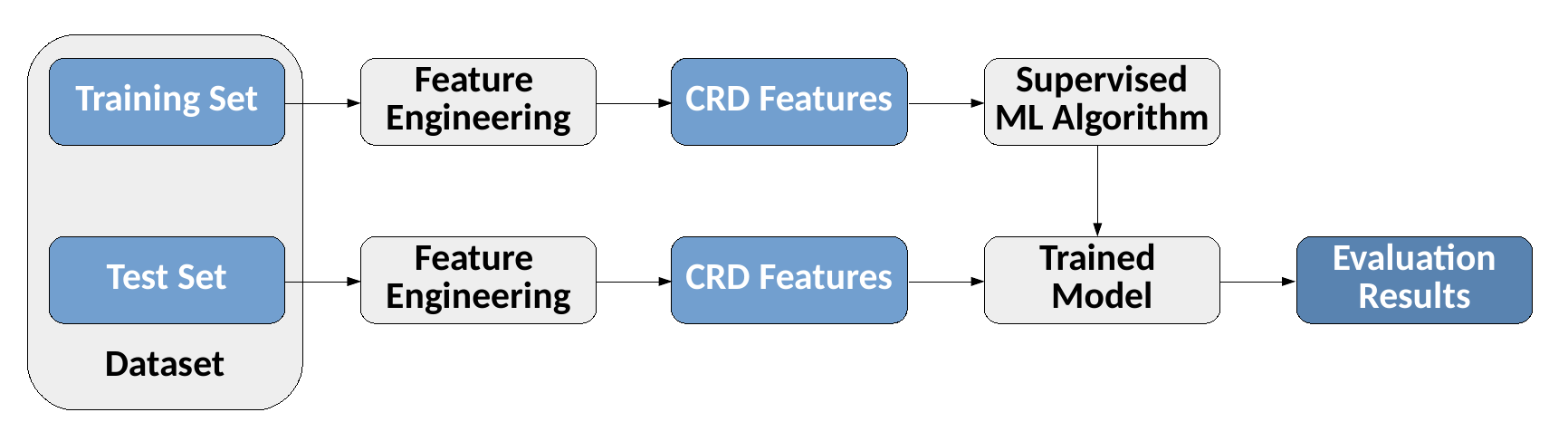}
    \caption{Experimental framework scheme.}\label{fig:framework}
\end{figure}

To ensure higher reliability of results, we apply a Stratified k-Fold Cross Validation approach~\citep{devijver1982pattern,kohavi1995astudy},
with $k=5$. 
The cross-validation involves partitioning the dataset into $k$ sets, then a model is trained using $k-1$ folds (called training set) and validated on the remaining part of the data (validation set). To reduce variability, this process is repeated $k$ times, using only once each partition for the validation. The performance measures are obtained by averaging the validation results on all runs. The Stratified approach ensures the preservation of the frequencies of the classes in each training and validation fold.
The experimental framework described so far is depicted in Figure~\ref{fig:framework}. All the experiments have been developed in Python, with the support of the Scikit-learn library~\citep{pedregosa2011scikit}.

\section{Evaluation and Discussion}
\label{sec:results}
In this section, we report the results obtained by applying the aforementioned algorithms for the detection of fake and genuine reviews, when using the basic or the cumulative features. Due to the imbalanced nature of the dataset, we report the performance metrics both for positive (fake-0) and negative (non fake-1) classes.

Table~\ref{tab:reviews_balanced_013_rating_simplef} and Table~\ref{tab:reviews_balanced_013_rating} report the results of the learning algorithms, obtained by using the basic and cumulative features, respectively. The performance values are computed by averaging the results over 5 runs of the algorithms, according to the employed cross-validation scheme. 

\begin{table}[hbt!]
    \centering
    \footnotesize
    \begin{tabular}{lrrrrrrrrrr}
    \toprule
        \textbf{Algorithm} & \textbf{TraAcc} & \textbf{TstAcc} & \textbf{Prec-0} & \textbf{Rec-0} & \textbf{F1-0} & \textbf{Prec-1} & \textbf{Rec-1} & \textbf{F1-1} & \textbf{MCC} &\textbf{AUC}   \\
        \midrule
        DT(Dep=10) & \textbf{0.99} & \textbf{0.97} & 0.77 & \textbf{1.00} & \textbf{0.87} & \textbf{1.00} & 0.96 & \textbf{0.98} & \textbf{0.86} & \textbf{0.98} \\ 
        
        DT(Dep=5) & 0.97 & 0.94 & 0.67 & 1.00 & 0.80 & 1.00 & 0.94 & 0.97 & 0.79 & 0.97 \\ 

        k-NN(k=5) & 0.93 & 0.92 & 0.71 & 0.79 & 0.75 & 0.96 & 0.94 & 0.95 & 0.70 & 0.95 \\
        
        k-NN(k=10) & 0.90 & 0.89 & 0.70 & 0.64 & 0.67 & 0.92 & 0.94 & 0.93 & 0.61 & 0.93 \\

        SVM(rbf) & 0.90 & 0.88 & \textbf{0.82} & 0.47 & 0.60 & 0.88 & \textbf{0.97} & 0.93 & 0.56 & 0.92 \\
       
        
        LogReg & 0.87 & 0.85 & 0.43 & 0.89 & 0.58 & 0.99 & 0.85 & 0.91 & 0.55 & 0.87 \\
        
        LDA & 0.84 & 0.81 & 0.36 & 0.87 & 0.51 & 0.98 & 0.80 & 0.88 & 0.48 & 0.84 \\

        Naive Bayes & 0.85 & 0.78 &	0.34 & 0.96 & 0.50 & 0.99 & 0.75 &	0.86 & 0.49 & 0.85 \\

        \bottomrule
    \end{tabular}
    \caption{Algorithms performances obtained using the Basic Features.}\label{tab:reviews_balanced_013_rating_simplef}
    \end{table}
    
\begin{table}[!t]
\centering
\footnotesize
    \begin{tabular}{lrrrrrrrrrr}
    \toprule
        \textbf{Algorithm} & \textbf{TraAcc} & \textbf{TstAcc} & \textbf{Prec-0} & \textbf{Rec-0} & \textbf{F1-0} & \textbf{Prec-1} & \textbf{Rec-1} & \textbf{F1-1} & \textbf{MCC} &\textbf{AUC}   \\
        \midrule
        DT(Dep=10) & \textbf{0.99} & \textbf{0.98} & \textbf{0.83} & \textbf{1.00} & \textbf{0.91} & \textbf{1.00} & \textbf{0.97} & \textbf{0.99} & \textbf{0.90} & \textbf{0.99} \\
        
        DT(Dep=5)  & 0.96 & 0.93 & 0.64 & 1.00 & 0.78 & 1.00 & 0.92 & 0.96 & 0.77 & 0.96 \\  
        
        k-NN(k=5) & 0.97 & 0.94 & 0.64 & 1.00 & 0.78 & 1.00 & 0.93 & 0.96 & 0.77 & 0.96 \\
        
        k-NN(k=10) & 0.95 & 0.91 & 0.56 & 1.00 & 0.72 & 1.00 & 0.90 & 0.95 & 0.71 & 0.95 \\
        
        SVM(rbf)     & 0.92 & 0.90 & 0.53 & 0.94 & 0.68 & 0.99 & 0.89 & 0.94 & 0.66 & 0.92 \\
       
        
        LogReg  & 0.91 & 0.89 & 0.45 & 0.93 & 0.66 & 0.99 & 0.89 & 0.93 & 0.64 & 0.91 \\
        
        LDA & 0.90 & 0.88 & 0.50 & 0.93 & 0.65 & 0.99 & 0.88 & 0.93 & 0.62 & 0.90 \\

        Naive Bayes & 0.86 & 0.86 & 0.44 & 0.86 & 0.58 & 0.98 & 0.86 & 0.91 & 0.55 & 0.86 \\

        \bottomrule
    \end{tabular}
    
    \caption{Algorithms performances obtained using the Cumulative Relative Frequency distributions.}\label{tab:reviews_balanced_013_rating}
    
\end{table}

Table~\ref{tab:reviews_balanced_013_rating_simplef} highlights that all the algorithms obtain very good precisions on the \textit{non-fake} class (tagged with 1), while the performances are worst when dealing with the \textit{fake} class. This is due to the fact that the dataset is imbalanced, with a ratio between classes of about 0.13. Among the experimented methods, the one that obtained the best precision on the minority class is the Support Vector Machine Classifier (precision = 0.97). However, the best global performances are obtained by the Decision Tree classifier (max depth=10), since in this case the MCC and the AUC reach the highest values (0.86 and 0.98, respectively). This occurs because the recall values obtained by the Decision Tree classifier are almost equal (majority class) or better (minority class) with respect to the ones obtained by the Support Vector Machine. We also experiment with a Decision Tree classifier with \textit{max depth=5}, to obtain a less complex model, but we notice that the performances on the minority class drop dramatically. Still regarding the precision on the minority class, all the remaining algorithms but the SVM are outperformed by Decision Tree (max depth=10). However, all of them, except the SVM, perform quite well with respect to the recall. This implies that the algorithms work well in finding the fake instances, but there are also many false positives, i.e., genuine reviews erroneously classified as fake. 

The results described so far and presented in Table~\ref{tab:reviews_balanced_013_rating_simplef} are generally worse with respect to those shown in Table~\ref{tab:reviews_balanced_013_rating}, where we reported the results obtained by using the Cumulative Relative Frequency distributions. Also when using the cumulative features, the best approach is the Decision Tree (max depth=10), which outperforms the other algorithms. By comparing the two tables, we notice that the results obtained by k-NN classifiers, using the basic features, reach higher values in the precision of the minority class, but this improvement in precision is balanced by a lower recall. 
In practice, the fake reviews detected by the classifiers are indeed fake, but they also miss a lot of actual fakes.

To summarize the performance differences, we report in Table~\ref{tab:comparison_simple_cum} a comparison between the MCC and the AUC values. In particular, for each metric, we reported the corresponding value obtained by using the basic features, the cumulative features, and the improvement percentage. Table~\ref{tab:comparison_simple_cum} highlights that the MCC and the AUC values are always higher or equal when using cumulative features, for all the algorithms except for the Decision Tree (max depth =5), since in this case the basic features lead to better MCC and AUC values.

\begin{table}[!bht]
\centering
\footnotesize
    \begin{tabular}{lrrrrrr}
    \toprule
        \textbf{Algorithm} & \multicolumn{3}{c}{MCC} & \multicolumn{3}{c}{AUC} \\
             & \textbf{Basic} & \textbf{Cumulative} & \textbf{Improv.(\%)} & \textbf{Basic} & \textbf{Cumulative} & \textbf{Improv.(\%)}   \\
    
        \midrule
        DT(Dep=10) & 0.86 & \textbf{0.90} &	4.7	& 0.98 &	 \textbf{0.99} & 1.0	\\
        DT(Dep=5) &	 \textbf{0.79} 	&	 0.77 	&	-2.5	&	 \textbf{0.97} 	&	 0.96 	&	-1.0	\\
        k-NN(k=5) &	 0.70 	&	 \textbf{0.77} 	&	10.0	&	 0.95 	&	 \textbf{0.96}	&	1.1	\\
        k-NN(k=10) &	 0.61 	&	 \textbf{0.71} 	&	16.4	& 0.93 	&	 \textbf{0.95}	&	2.2	\\
        SVM(rbf) & 0.56	& \textbf{0.66}	&	17.9	&	 \textbf{0.92} 	&	 \textbf{0.92}	&	0.0	\\
        LogReg  &  0.55 	&	 \textbf{0.64} 	&	16.4	&	 0.87	&	 \textbf{0.91}	&	4.6	\\
        LDA	&	 0.48 	&	 \textbf{0.62} 	&	29.2	& 0.84 	&	 \textbf{0.90}	&	7.1	\\
        Naive Bayes	&	 0.49 	&	 \textbf{0.55} 	&	12.2	& 0.85 	&	 \textbf{0.86}	&	1.2	\\
        \bottomrule
    \end{tabular}
    \caption{Performances comparison between Basic and Cumulative Features.}\label{tab:comparison_simple_cum}
    \end{table}

We finally remark that the process described in Section~\ref{sec:engineering} represents a pre-processing step and it can be performed only once before applying several machine learning algorithms. We computed the time required to build the Cumulative Relative Frequency for all the features in the pre-processing of the cross-validation process and we obtained an average value of 155 ms, pointing out that the impact of the  calculation time for the proposed features is negligible.

\subsection{Features Importance}
\begin{figure}
    \centering
    \includegraphics[width=0.7\textwidth]{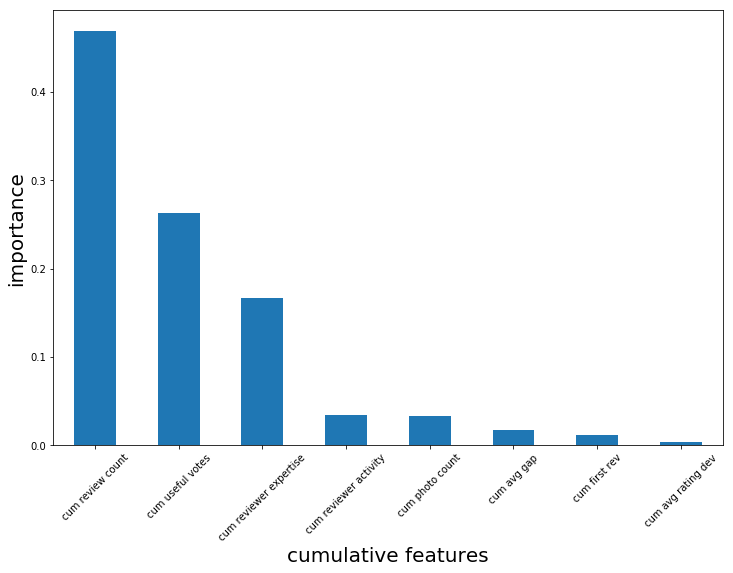}
    \caption{Feature importance of cumulative features computed according to Gini index, for the Decision Tree model (max depth = 10). }
    \label{fig:significance}
\end{figure}

The importance of features plays a significant role in a predictive model, since it provides insights into the data and into the model itself. Moreover, it poses the basis for dimensionality reduction and feature selection, which can sometimes improve the efficiency and the effectiveness of a predictive model.
From the experiments carried on, the best algorithm resulted to be a Decision Tree model. Decision Tree algorithms offer importance scores based on the reduction in the criterion used to select split points in the tree, such as Gini or entropy. In this case, the importance of a feature has been computed following the Gini importance ~\cite{breiman1984classification}, which counts the times a feature is used to split a node, weighted by the number of samples it splits. We computed the Gini importance in each fold of the cross-validation and we averaged the results obtained. In Figure \ref{fig:significance}, we report a graph showing the importance of each cumulative feature, from the most important to the least one.
The first three features, namely the number of reviews posted on the platform (\textit{cum review count}), the number of useful votes received (\textit{cum useful votes}), and the membership length of a reviewer (\textit{cum reviewer expertise}) are probably the most useful for this predictive model.

To confirm this intuition, we perform a feature selection process, by selecting the first three most important features. A new Decision Tree model has been trained and tested, and the results (averaged on 5-folds) are reported in Table~\ref{tab:feat_selection_results}. The features selected are the same for both models, although they appear in different order of importance. For the sake of clarity, we reported in Table~\ref{tab:feat_selection_results} the results already presented in Table~\ref{tab:reviews_balanced_013_rating}, related to the Decision Tree models without feature selection. From their comparison, we notice that the feature selection process actually contributes to slightly improving the performances. In particular, for the Decision Tree model with \textit{max depth=5}, this improvement is more evident since all the metrics reach higher values when the feature selection is applied. On the other hand, for the Decision Tree model with \textit{max depth=10}, there is a small improvement in the precision and the F1-score of the positive (fake-0) class, whereas the recall for the negative (non fake-1) class is a bit lower with the feature selection applied. Nevertheless, the global performances are better, since the MCC value is higher.

\begin{table}[!bth]
\centering
\footnotesize
    \begin{tabular}{llrrrrrrrrrr}
    \toprule
        \textbf{Algorithm} & \textbf{Selected Features} & \textbf{TraAcc} & \textbf{TstAcc} & \textbf{Prec-0} & \textbf{Rec-0} & \textbf{F1-0} & \textbf{Prec-1} & \textbf{Rec-1} & \textbf{F1-1} & \textbf{MCC} &\textbf{AUC}   \\
        \midrule
        DT(Dep=10) & c.rev.count, c.useful votes, c.rev.exp. & \textbf{0.99} & \textbf{0.98} & \textbf{0.85} & \textbf{1.00} & \textbf{0.92} & \textbf{1.00} & \textbf{0.98} & \textbf{0.99} & \textbf{0.92} & \textbf{0.99} \\

        DT(Dep=5) & c.rev.count, c.rev.exp., c.useful votes  & 0.97 & 0.95 & 0.70 & 1.00 & 0.82 & 1.00 & 0.94 & 0.97 & 0.81 & 0.97 \\  
        \midrule
        DT(Dep=10) & all & 0.99 & 0.98 & 0.83 & 1.00 & 0.91 & 1.00 & 0.97 & 0.99 & 0.90 & 0.99 \\
        
        DT(Dep=5) & all  & 0.96 & 0.93 & 0.64 & 1.00 & 0.78 & 1.00 & 0.92 & 0.96 & 0.77 & 0.96 \\  
        \bottomrule
    \end{tabular}
    
    \caption{Algorithms performances obtained with and without feature selection for the Decision Tree models.}\label{tab:feat_selection_results}
    
\end{table}
\subsection{Discussion on Feature Enhancement}
In this section, we explain at a conceptual level why using the cumulative relative frequencies works better than the simple values. 
In data pre-processing, the normalization task changes the feature values to a common scale to avoid that differences in the data ranges can distort the results. Min-max normalization is one of the most used methods and consists in re-scaling the values of features in the range [0, 1]. However, min-max normalization suffers from outliers, because it is sufficient one relevant outlier to flatten all the input values. Another normalization method that transforms the feature values in such a way that they have zero-mean is Z-score. This method avoids the outlier issue but does not produce data on the same scale.

Several probabilities distributions are affected by the presence of outliers. One of them is the Zipfian distribution, which is a discrete power-law probability distribution \cite{zipf1949human} in which the frequency of input is inversely proportional to its rank in the frequency table. That is, the most frequent inputs occur twice as often as the second most frequent ones, three times as often as the third most frequent inputs, and so on. For example, consider the number of friends in a social network: normally, most users have very few friends (usually, 0 is the most frequent value), whereas very few users have a very high number of friends. In these cases, the performance of min-max normalization gets worse due to outliers, whereas Z-score does not well scale the input data.

Our approach allows us to scale the value of a feature in the desired range [0, 1] using its rank in the frequency table, in a similar way as the Zipfian distribution. This transformation does not suffer from outliers: indeed, the normalized value of a feature does not depend on the magnitude of an outlier. Moreover, the normalized values are better distributed.

Consider again the example of the number of friends in a social network presented above: the normalized feature $v$ of a user with no friend is equal to zero using any standard normalization method. In our approach, $v=\frac{n_0}{n}$, which is the fraction of users with no friends.

By examining the features used in our experiments, we found that the most important ones follow a Zipfian distribution. This result confirms the findings of other research works, such as \cite{muchnik2013origins,adamic2001search,raban2009statistical,arenas2004community}, that observed a power-law distribution in many social network dimensions. 

We expect that our strategy performs well in scenarios in which input data are distributed according to a power-law probability distribution because our feature normalization is strongly derived from the power-law distribution. Thus, as a practical recommendation in a classification task, we suggest to identify the probability distribution of each feature: in the case such a distribution follows the power law, then re-engineering this feature by considering the Cumulative Relative Frequency Distribution should improve the overall accuracy of the classifier performances. By considering that, in many fields, such as physical and social sciences \cite{clauset2009power}, data follow a Zipfian distribution, we conclude that our normalization strategy can be effective in many contexts.

\section{Conclusions}
\label{sec:concl}
User opinions 
are an important information source, which can help a customer and a vendor to evaluate pros and cons of the buying/selling when they interact. For the importance of opinion role, there is the possibility to have unfair opinions used to promote own products or to disparage products of competitors. The important challenge of detecting unfair opinions has attracted and attracts the scientific community and one of the most promising approaches to address this problem is based on the use of supervised classifiers, which have been proven to be highly effective. 

In this paper, we tried to further improve their effectiveness, not by proposing some change in the well-tested state-of-the-art algorithms, but only by modifying the input used for the training phase to construct supervised classifiers. Specifically, we considered eight features widely used to detect opinion spam and pre-processed them by considering the cumulative relative frequency distribution. To demonstrate the effectiveness of our proposal, we extracted a data set from Yelp.com and measured the performances of the six most used classifiers in detecting opinion spam, both in their standard use and when our proposal is adopted. The results of this comparison show that the use of the cumulative relative frequency distribution improves the performance of the state-of-the-art classifiers.

As future work, we intend to extend our proposal to detect not only individual spammers, 
but also groups of users who, acting in a coordinated and synchronized way, aim to give credit or discredit a product (or a service). The idea is that, once an ensemble of malicious reviewers is detected, an overlapping between the products that malicious reviewers have evaluated is searched. Groups of users with large overlap (i.e., who revised the same products) could be colluders.

\begin{acks}
Partially supported by the European Union's Horizon 2020 programme (grant agreement No. 830892, SPARTA); by the Integrated Activity Project TOFFEe (TOols for Fighting FakEs), funded by IMT Scuola Alti Studi Lucca; and by the IIT-CNR project DESIRE (DissEmination of ScIentific REsults).
\end{acks}

\bibliographystyle{ACM-Reference-Format}
\bibliography{biblio}


\end{document}